\documentstyle{ichep}
\def\simge{\mathrel{%
   \rlap{\raise 0.511ex \hbox{$>$}}{\lower 0.511ex \hbox{$\sim$}}}}
\def\simle{\mathrel{
   \rlap{\raise 0.511ex \hbox{$<$}}{\lower 0.511ex \hbox{$\sim$}}}}

\def\slashchar#1{\setbox0=\hbox{$#1$}           
   \dimen0=\wd0                                 
   \setbox1=\hbox{/} \dimen1=\wd1               
   \ifdim\dimen0>\dimen1                        
      \rlap{\hbox to \dimen0{\hfil/\hfil}}      
      #1                                        
   \else                                        
      \rlap{\hbox to \dimen1{\hfil$#1$\hfil}}   
      /                                         
   \fi}                                         %

\def\ol{\overline}

\def\ecm{\sqrt{s}}
\def\gev{\rm GeV}
\def\tev{\rm TeV}
\def\ts{\thinspace}

\def\ra{\rightarrow}

\def\bbb{b \ol b}
\def\Dbb{\Delta_{b \ol b}}

\def\pt{p_T}

\def\atc{\alpha_{TC}}

\def\cond{\langle \overline\psi \psi\rangle}

\def\gmm{\gamma_m}

\begin{document}

\title{Technicolor and Precision Tests of the Electroweak
Interactions$^{\dag}$}

\author{Kenneth Lane$^{\ddag}$}

\affil{Department of Physics,
Boston University,\\
590 Commonwealth Avenue, Boston, MA 02215, USA\\}

\abstract{Precision electroweak measurements have been claimed to eliminate
almost all models of technicolor. We show that the assumptions made to
calculate the oblique parameters $S$,$T$,$U$ apply to QCD--like technicolor
models which were ruled out long ago on much firmer grounds. These
assumptions are invalid in modern ``walking'' technicolor models.}

\twocolumn[\maketitle]

\fnm{1}{Invited talk given at the 27th International Conference on High
Energy Physics, Glasgow, 20--27th July 1994.}
\fnm{2}{email: lane@buphyc.bu.edu}

\section{Introduction}

Technicolor---dynamical breaking of electroweak symmetry by an
asymptotically free gauge interaction---was invented in
1979~\cite{swtc},\cite{lstc}. Extended technicolor---the gauge mechanism
for introducing quark and lepton flavors and for breaking flavor
symmetry---followed quickly~\cite{dsetc},\cite{eekletc}. It was already
pointed out in Ref.~\cite{eekletc} that ETC theories generically have large
flavor--changing neutral currents and that an ETC scale $\Lambda_{ETC}$ of
$O(1000\,\tev /times \ts {\rm mixing \ts\ts angles})$ was needed to avoid
conflict with experiment in the neutral kaon system (also see
Ref.~\cite{ellis}). This large scale implied ridiculously small quark and
lepton masses, as well as light pseudo--Goldstone bosons (technipions) that
were soon ruled out by experiment. Technicolor was dead.

The most promising solution to the FCNC problem was not long in
coming~\cite{holdom}. Holdom showed that, if the technifermion bilinear
condensate, $\cond$, has a large anomalous dimension, $\gmm$, it is
possible to have a very large ETC scale without unduly small fermion
and technipion masses. Unfortunately, Holdom did not provide a convincing
field--theoretic explanation of how this large $\gmm$ could occur. His idea
lay dormant until 1986 when several groups pointed out that a technicolor
gauge theory with a very slowly running coupling, $\atc(\mu)
\simeq$~constant for $\Lambda_{TC} \sim 1\,\tev < \mu < \Lambda_{ETC}$,
gives rise to
$\gmm(\mu) \simeq 1$ over this large energy range~\cite{wtc}. This
``walking technicolor'' permitted the increase in $\Lambda_{ETC}$ needed to
eliminate FCNC. Thus, the resurrection of technicolor was brought about by
abandoning the notion that its gauge dynamics were QCD--like, with {\it
precocious} asymptotic freedom and all that implies. (For a recent review
of technicolor, its problems and proposed solutions, see
Ref.~\cite{kltasi}.)

In 1990, it was rediscovered that technicolor dynamics (TC, {\it not} ETC,
in this case) could affect electroweak parameters that were just then
beginning to be very precisely measured in LEP experiments~\cite{pettests}.
Estimating the effects of technicolor on the ``oblique'' parameter $S$ (or
its equivalent), many authors showed that one--family TC models were
inconsistent with its then--measured value. Once again, technicolor was
dead.

This news received considerable attention in journals and on the conference
circuit. Little attention was given to the protests of technicolor
afficionados that the technicolor killed by the precision tests had been
dead for a decade. Walking technicolor was {\it not} ruled out by these
tests and it remains unclear how to confront it with the precision
electroweak measurements. I am grateful to the organizers of the ``Beyond
the Standard Model'' session at this conference for this opportunity to
review the issues. I will do that as clearly as I can.

In the next section, I state the definition of the $S$,$T$,$U$
parameters popularized by Peskin and Takeuchi, give their most recent
values, and detail the assumptions that have been used to calculate these
parameters in technicolor models. In Section~3, I show that all these
assumptions are wrong or, at best, questionable in walking technicolor.
This, of course, will not convince its detractors that technicolor is still
viable; nor is it intended to. My intent is to persuade that the
$S$,$T$,$U$--argument against technicolor is far from made. Finally, in
Section~4, I discuss some other aspects of precision electroweak tests.
These include the question of technicolor (here ETC is involved) and the
rate for $Z^0 \ra \bbb$, as well as some other curiosities in the precision
electroweak data.

\section{Technicolor and Precision Electroweak Tests---The Problem}

The standard $SU(2) \otimes U(1)$ model of electroweak interactions has
passed all experimental tests faced so far. The parameters of this
model---$\alpha(M_Z)$, $M_Z$, $\sin^2 \theta_W$---are so precisely known
that they may be used to limit new physics at energy scales above 100~GeV.
The quantities most sensitive to new physics are defined in terms of
electroweak current correlation functions:
\begin{equation}\eqalign{
&\int d^4x \ts e^{-i q\cdot x} \langle\Omega | T\bigl(j^\mu_i(x)
j^\nu_j(0)\bigr) | \Omega \rangle =\cr
&\qquad\qquad i g^{\mu\nu} \Pi_{ij}(q^2) + q^\mu q^\nu \ts {\rm terms}
\ts.\cr}
\end{equation}
Assuming that the scale $\Lambda_{\rm new}$ of this physics is {\it well
above} $M_{W,Z}$, one may define ``oblique'' correction factors $S$, $T$
and $U$ that measure its effects by
\begin{equation}\eqalign{
S&= 16\pi \bigl[ \Pi_{33}^{'}(0) - \Pi_{3Q}^{'}(0)
\bigr] \ts , \cr
T&= {4\pi \over{M^2_Z \cos^2\theta_W \sin^2\theta_W}}
\ts \bigl[ \Pi_{11}(0) - \Pi_{33}(0) \bigr] \ts , \\
U&= 16\pi \bigl[ \Pi_{11}^{'}(0) - \Pi_{33}^{'}(0)
\bigr] \ts . \cr}
\end{equation}
Here, the prime denotes differentiation at $q^2 = 0$, and these are the
leading terms in an expansion in $M^2_Z/\Lambda^2_{\rm new}$.
The parameter $S$ is a measure of the splitting between $M_W$ and $M_Z$
induced by weak--isospin conserving effects. The parameter $T$ is defined
in terms of $\rho \equiv M_W^2/M_Z^2 \cos\theta_W^2 = 1 + \alpha T$. The
$U$--parameter measures weak--isospin breaking in the $W$ and $Z$ mass
splitting.

Langacker recently made a global ``best fit'' to a set of
precisely--measured electroweak quantities, using the CDF measurement of
the top--quark mass, $m_t = 174\pm 10 \ts ^{+13}_{-12} \,\gev$\cite{cdf}.
He extracted the following values of $S,T,U$ due to potential {\it new
physics}~\cite{pgl}:
\begin{equation}\eqalign{
S &= -0.15 \pm 0.25^{-0.08}_{+0.17} \cr
T &= -0.08 \pm 0.32^{+0.18}_{-0.11} \\
U &= -0.56 \pm 0.61 \ts .\cr}
\end{equation}
The first error is the net experimental error, assuming a standard Higgs boson
mass of 300~GeV; the second error is the effect of varying $M_H$ from 60
to 1000~GeV.

It is clear that $S,T,U$ can be computed precisely only if the new physics
is weakly coupled. It would have been impossible to calculate the QCD
analogs of $S,T,U$ without experimental information on the hadronic weak
currents---the color and flavor of quarks, the spectrum of hadrons, and so
on. New standard--model data is still leading to revisions. A year ago, the
quoted value of $S$ was rather different, $-0.8\pm 0.5$, from Eq.~(3) (see
\cite{kltasi}); the change is due to the fact that we now know the
top--quark mass~\cite{cdf}. No data is available for technicolor---a strong
gauge theory at a scale of several 100~GeV. Assumptions must be made to
estimate its contributions to $S,T,U$.

The assumptions made to calculate $S$ amount to assuming that technicolor
is just QCD scaled up to a higher energy, with $N_D$ elecroweak doublets of
technifermions belonging to the fundamental representation of a strong
$SU(N_{TC})$ technicolor gauge group:

\medskip

\noindent 1.) Techni-isospin is a good symmetry, i.e., custodial $SU(2)$
breaking by ETC interactions is negligible.

\noindent 2.) Asymptotic freedom sets in quickly above the technicolor
scale $\Lambda_{TC}$.

\noindent 3.) Appropriate combinations of spectral functions of current
correlators may be estimated using vector--meson dominance, i.e.,
saturating the spectral integrals with the lowest--lying spin--one
resonances. Why this works in QCD is a mystery, but it is consistent with the
precocious asymptotic freedom of QCD (see the discussion in Section~3).

\noindent 4.) The spectrum of techni-hadrons may be scaled from QCD using,
e.g., large--$N_{TC}$ arguments.

\noindent 5.) Chiral lagrangians may be used to describe the low--energy
dynamics of technipions, with coefficients of terms scaled from the
QCD values\cite{gass}.

\medskip

As an oft--cited example of how these assumptions are employed, I present a
simplified version of Peskin and Takeuchi's calculation of $S$~\cite{pettests}.
If techni--isospin is a good symmetry, then $S$ may be written as the
following spectral integral:
\begin{equation}\eqalign{
S &= -4 \pi \left[ \Pi_{VV}^{'}(0) - \Pi_{AA}^{'}(0) \right]\cr
&= {1 \over {3\pi}} \int_0^\infty {ds \over {s}} \left [ R_V(s) - R_A(s)
\right] \ts. \cr}
\end{equation}
Here, $\Pi_{VV(AA)}$ is the polarization function for the product of two
vector (axial-vector) weak isospin currents (e.g., $j^3_\mu j^3_\nu$);
$R_{V(A)}$ is the analog for these current of $R(s) =
\sigma(e^+e^- \ra {\rm hadrons}) / \sigma(e^+e^- \ra \mu^+ \mu^-)$.
They are the spin--one spectral functions to which Weinberg's two sum
rules apply~\cite{sfsr}:
\begin{equation}\eqalign{
& \int_0^\infty ds \left [ R_V(s) - R_A(s) \right] = F_\pi^2 \cr
& \int_0^\infty ds \cdot s \left [ R_V(s) - R_A(s) \right] = 0 \ts .\cr}
\end{equation}
These sum rules, written here for conserved currents, are implied by the
strength of the singularity at $x \ra 0$ in $\langle\Omega | T(j_{L\mu}(x)
j_{R\nu}(0)) |\Omega\rangle$. The second sum rule, in particular, requires
asymptotic freedom for its validity. In Eqs.~(5), $F_\pi = 246\,\gev$ is
the decay constant of the technipions that become the longitudinal
components of the $W$ and $Z$.

In the evaluation of $S$, the spectral functions
$R_V$ and $R_A$ are approximated by saturating them with the
lowest--lying vector ($\rho_T$) and axial--vector ($a_{1T}$) meson poles,
using Eqs.~(5) to help fix their parameters. Their masses are
scaled from QCD using large-$N_{TC}$. In the narrow width approximation,
\begin{equation}
S = 4 \pi \left(1 + {M^2_{\rho_T} \over{M^2_{a_{1T}}}}\right ) {F^2_\pi \over
{M^2_{\rho_T}}} \simeq 0.25 \ts N_D \ts {N_{TC}\over{3}} \ts.
\end{equation}
It appears from Eq.~(6) that all technicolor models with $N_D > 1$
and $N_{TC} > 3$ are ruled out; this includes the popular one--family model
($N_D = 4$).

The other main method of calculating $S$ uses chiral lagrangians.
Technicolor models with $N_D$--doublets have $4(N_D^2 -1)$ physical
technipions. Golden and Randall, Holdom and Terning, and
others~\cite{pettests} estimated their leading chiral--logarithmic
contribution, $S_{\pi_T}$, to $S$$^{\dag}$\fnm{1}{Holdom and Terning also
estimated the non--chiral log part of the relevant coefficient in the
chiral lagrangian by scaling from QCD. I will discuss below why such
scaling is problematic.}. This approach is valid, independent of the nature
of technicolor dynamics, so long as ETC interactions are weak enough that a
chiral perturbation expansion is accurate. Assuming all technipions are
degenerate and that the cutoff scale for the chiral logs is $M_{\rho_T}$,
these authors obtained
\begin{equation}
S > S_{\pi_T} \simeq {1 \over {12 \pi}} (N_D^2 -1) \log\left({M^2_{\rho_T}
\over{M^2_{\pi_T}}} \right)\simeq 0.08 (N_D^2-1) \ts.
\end{equation}
Eqs.~(6) and (7) agree for the popular choice of the
one--family model, $N_D=N_{TC}=4$, in which case $S \simeq 1$, almost
$4\sigma$ away from the central value quoted above. This agreement is
accidental; see Ref.~\cite{cdg}. Nevertheless, except for the simplest
possible technicolor model, such estimates of $S$ have led to the
oft--repeated observation that, to paraphrase Ref.~\cite{Oz}, ``technicolor
is not only really very dead, it's really most sincerely dead!''

\section{Walking Technicolor and {\bf $S$,$T$,$U$}}

While chiral symmetry breaking and bound state formation in QCD are
nonperturbative phenomena, requiring strong--coupling methods for their
study, much interesting physics of quarks and gluons occurs above 1~GeV
where it is possible to exploit asymptotic freedom. Walking technicolor, is
essentially nonperturbative over the entire range, $\Lambda_{TC}$ to
$\Lambda_{ETC}$. Let us see how this affects the basic assumptions made in
calculating $S$. For now, leave the question of techni--isospin aside. That
has as much to do with ETC interactions as with walking TC.

The assumption that asymptotic freedom sets in quickly above $\Lambda_{TC}$
is patently wrong. This assumption was used implicitly (and is essential)
in approximating the spectral functions $R_V(s)$ and $R_A(s)$. It tells how
these functions behave at large~$s$ and, in turn, how fast $\Pi_{VV}(q^2) -
\Pi_{AA}(q^2)$ falls at large $q^2$. In an asymptotically free theory,
$\Pi_{VV}(q^2) - \Pi_{AA}(q^2) \sim q^{-4}$ above $\Lambda_{TC}$. In a
walking gauge theory, $\Pi_{VV}(q^2) - \Pi_{AA}(q^2) \sim q^{-2}$ until the
coupling becomes small, at $q^2 \simle \Lambda_{ETC}^2$. Consequently, the
convergence of the second spectral integral to zero (Eq.~(5)) is much
slower in a walking gauge theory, and $R_V - R_A$ cannot be approximated by
a {\it single, close pair} of vector and axial--vector meson poles. It
follows that the masses and widths of hadrons in a walking gauge theory
cannot simply be scaled up from QCD; the spectrum of a walking gauge theory
is a mystery. While the integral for $S$ is dominated by low energies, the
spectral sum rules connect the low and high energy behavior of $R_V - R_A$.
In a walking theory, the spectral weight of $R_V - R_A$ is shifted to
higher energies. Thus, it is possible that $S$ is smaller in such a theory
than in a QCD--like one.

Another reason to be skeptical of scaling from QCD is that some or all
technifermions may belong to higher-dimensional representations of the TC
gauge group. Then, large--$N_{TC}$ arguments are inapplicable.

The assumption of a reliable chiral--perturbative expansion in a walking
gauge theory is also unjustified. Like the technifermion bilinear, the
operators $\ol\psi\psi\ol\psi\psi$ involved in ETC generation of technipion
masses have large anomalous dimensions~\cite{wtc}. In the extreme walking
case, these become relevant operators so that $M_{\pi_T} \sim
\Lambda_{TC}$; i.e., the technipions are {\it not} approximate Goldstone
bosons. In generic walking TC theories, then, the chiral Lagrangian
estimate of a lower bound for $S$ is also likely to be incorrect.

Now return to the question of techni--isospin conservation, and $T$ and $U$
as well. This assumption appeared plausible because, otherwise, $T$ ought
to be too large. However, ETC theories need to have rather large isospin
breaking to account for the top--quark mass of $O(F_\pi)$! Can this be
consistent with small $S$ and $T$?

The $T$--parameter is notoriously difficult to calculate (which partly
explains why so few attempt it). The main problem is that $T$ is directly
determined by physics at higher scales ($\Lambda_{TC}$) even in ordinary TC
theories; there is no derivative in its definition (see, e.g.,
Ref.~\cite{tatcalc}). This may point the way out. It is possible that there
are several scales of chiral symmetry breaking in TC theories (e.g., see
Ref.~\cite{multi})$^{\dag}$\fnm{1}{If the lower scales in these multiscale
TC models are close to $M_Z$, the assumption that oblique corrections are
characterized by just the lowest derivatives $S$,$T$,$U$ is also
incorrect.}. The highest scales, mainly responsible for generating
$M_{W,Z}$, may respect weak isospin. The lower scales, which contribute to
$S$, may not. It has long been known and was emphasized in
Ref.~\cite{tajtstu} that this can lead to a small and even a negative value
for $S$. Whether multiscale theories can generate a large~$m_t$ is a
model--dependent question. See Ref.~\cite{tajtmod} for an example that may
produce large $m_t$. There is practically nothing we can say about $U$. It
is generally presumed to be of $O(S \cdot T)$. We are unaware of attempts
to compute it in a walking technicolor theory.

\section{Other Electroweak Discrepancies}

The deviation of the measured $Z^0 \ra \bbb$ rate from the standard--model
expectation is~\cite{pgl}
\begin{equation}
\Dbb \equiv \Gamma(Z \ra b \ol b)/\Gamma(Z \ra b \ol b)_{SM} -1
=  0.031 \pm 0.014 \ts,
\end{equation}
i.e., $2.2\sigma$ away from zero. This rate may turn out to be the most
incisive test of TC/ETC theories. The reason for this is that the
top--quark is so heavy that the ETC boson which generates $m_t$ is probably
very light, of order a few~TeV (an exception to this will be mentioned
below). Consequently, the Fierzed ETC interaction
\begin{equation}
\xi^2\frac{1}{%
\Lambda^2_{ETC}(t)}\left( \ol{T}_L\gamma ^\mu \frac{\overrightarrow{\tau }}%
2T_L\right) \cdot \left( \ol{\psi }_L\gamma _\mu \frac{\overrightarrow{\tau }%
}2\psi _L\right)
\end{equation}
modifies the coupling of left--handed $b$--quarks to the
$Z^0$\cite{rsc}. Here, $\Lambda_{ETC}(t)$ is the ETC scale involved in
generating $m_t$; $T_L$ is a left--handed technifermion doublet and $\psi_L =
(t,b)_L$; and $\xi$ is a model--dependent factor expected to be $O(1)$.

In a QCD--like technicolor theory,
\begin{equation}
\Dbb = -0.065 \xi^2 \left({m_t \over{175\,\gev}}\right) \ts,
\end{equation}
in clear conflict with the value quoted in Eq.~(8). The situation is
little improved if $\atc$ walks because a low $\Lambda_{ETC}(t)$
is still needed to produce such a large $m_t$~\cite{gates}.

Clearly this is a problem of the ETC, not just the TC, interaction.
Two modifications to ETC can eliminate the conflict with $\Dbb$. The
first, which appears to be necessary anyway to explain the large $m_t$, is
known as strong extended technicolor (SETC). An ETC scale of
$O(1\,\tev)$ makes no sense dynamically. There is not enough splitting between
the scale at which ETC breaks to TC and the TC scale
itself. To maintain a substantial hierarchy between $\Lambda_{ETC}(t)$ and
$\Lambda_{TC}$, it seems necessary that some ETC interactions
be strong enough to participate with TC in the breakdown of electroweak
symmetry~\cite{setc}. This requires some fine tuning of the ETC coupling and
leads to a composite scalar state light compared to
$\Lambda_{ETC}(t)$~\cite{ccl}. The increased $\Lambda_{ETC}(t)$ leads to a
$\Dbb$ too small to detect~\cite{evans}.

The second modification of ETC which can eliminate conflict with $\Dbb$ is to
give up the time--honored, but apparently inessential, assumption that the
ETC gauge group commutes with electroweak $SU(2)$$^{\ddag}$\fnm{2}{See
\cite{eekletc} for a discussion of this assumption.}. Chivukula, Simmons and
Terning have recently considered the magnitude of $\Dbb$ in such noncommuting
ETC theories {\it without} the assumption of SETC~\cite{liz}. They found that
it is possible to obtain $\Dbb$ of order the value in Eq.~(10), but with {\it
either} sign. This will be especially interesting if the deviation in Eq.~(8)
survives.

Finally, I draw attention to two other curiosities in the precision
measurements. The first involves $\sin^2\theta_W$. The SLD measurement
reported at this conference is~\cite{fero}
\begin{equation}
\sin^2\theta_W({\rm SLD}) = 0.2292 \pm 0.0009 \pm 0.0004 \ts.
\end{equation}
The LEP average value reported here is~\cite{moenig}
\begin{equation}
\sin^2\theta_W({\rm LEP}) = 0.2321 \pm 0.0003 \pm 0.0004 \ts.
\end{equation}
These differ by $2.9\sigma$. An equivalent (and perhaps more direct)
expression of this intercontinental disagreement is provided by the
left--right asymmetry. The SLD measurement is (from Ref.~\cite{pgl}, whose
notation we follow)
\begin{equation}
A^0_e({\rm SLD}) = 0.164 \pm 0.008
\end{equation}
The asymmetry inferred from LEP measurements of the forward--backward
asymmetry in $e^+e^- \ra Z^0 \ra e^+e^-$ and the angular distribution of
$\tau$--polarization is
\begin{equation}
A^0_e({\rm LEP}) = 0.129 \pm 0.010
\end{equation}
The disagreement here is $2.7\sigma$.

The second discrepancy is smaller and wouldn't be worth mentioning if it
weren't in a quantity of such great theoretical interest. It is the QCD
coupling renormalized at $M_Z$, $\alpha_S(M_Z)$. The LEP average value,
extracted from the $Z^0$ lineshape, is~\cite{pgl}
\begin{equation}
\alpha_S(M_Z | {\rm LEP}) = 0.124 \pm 0.005 \pm 0.002 \ts.
\end{equation}
Most low--energy measurements of $\alpha_S(M_Z)$ give a lower value. The one
with the smallest quoted error is extracted from the charmonium spectrum using
lattice--QCD methods to separate out the confining potential's
contribution~\cite{lattice}:
\begin{equation}
\alpha_S(M_Z | {\rm Lattice}) = 0.115 \pm 0.002 \ts.
\end{equation}
These values differ by $1.5\sigma$. Langacker stresses that the value of
$\alpha_S(M_Z)$ extracted from the $Z^0$ lineshape is sensitive to certain
types of new physics. His global fit, allowing a nonzero $\Dbb$, gave the
result in Eq.~(8) {\it and} $\alpha_S(M_Z) = 0.103 \pm 0.11$, $2\sigma$
away from the LEP measurement.

What are we to make of these discrepancies? The deviation $\Dbb$ is
$2\sigma$ from zero. Shall we say that the standard model is ruled out?
Surely, almost everyone believes that will happen someday. The LEP and SLD
measurements of $\sin^2\theta_W$ differ by almost $3\sigma$. Is this just
experimental error? If so, who's wrong? Low and high--energy determinations
of the QCD coupling are on the verge of being inconsistent. Is this
just~(!) the effect of new physics at high energies? Given these
discrepancies, might it not be premature to say that essentially
nonperturbative theories such as walking technicolor are ruled out by the
values of $S$ and $T$. At the very least, we ought to bear in mind Vernon
Hughes' admonition~\cite{taubes}:

\bigskip

\centerline{\bf {Half of all three sigma measurements are wrong.}}

\bigskip

I am grateful to Sekhar Chivukula, Mitchell Golden, Elizabeth Simmons and John
Terning for their careful reading of the manuscript and helpful comments.
This research was supported in part by the Department of
Energy under Contract~No.~DE--FG02--91ER40676.

\end{document}